\documentclass{article}

\usepackage{cite}
\usepackage{graphicx}
\usepackage{dcolumn}

\begin{document}

\title{Improved power-series method for confined one-dimensional quantum-mechanical
problems}
\author{Francisco M. Fern\'{a}ndez\thanks{fernande@quimica.unlp.edu.ar} \\
INIFTA (UNLP, CCT La Plata-CONICET), Divisi\'on Qu\'imica Te\'orica,\\
Blvd. 113 S/N, Sucursal 4, Casilla de Correo 16,\\
1900 La Plata, Argentina}
\maketitle

\begin{abstract}
We propose two improvements to the well-known power series method for
confined one-dimensional quantum-mechanical problems. They consist of the
addition of a variational step were the energy plays the role of a
variational parameter. We compare the rate of convergence of the three
methods on an exactly-solvable model. We also outline possible
generalizations of the approaches to more complex problems.
\end{abstract}

\section{Introduction}

\label{sec:intro}

Confined quantum-mechanical models have proved suitable for the study of
several physical phenomena and there is a vast literature on the subject\cite
{AQC09}. Here we are interested in a power-series approach that is suitable
for obtaining accurate solutions to simple models. As an example, Patil and
Varshni\cite{PV09} applied it to the hydrogen atom in a spherical box with
impenetrable walls. The approach consists of calculating a finite number of
terms of the Taylor series of the solution about the center of the box at $%
r=0$ and then requiring that the solution satisfies the boundary condition
on the wall at $r=R$. From the resulting equation one obtains an approximate
energy that becomes increasingly accurate as the number of terms in the
power series increases.

The purpose of this paper is to propose two possible improvements to the
power-series approach outlined above. In section~\ref{sec:model} we briefly
discuss the Schr\"{o}dinger equation for a sufficiently general
one-dimensional confined model. In section~\ref{sec:approaches} we develop
the main ideas of those methods. In section~\ref{sec:results} we choose a
benchmark model for the comparative application of the approaches. The
chosen model can be solved exactly and is therefore useful for monitoring
the rate of convergence of the methods because none of them yields the exact
analytical result. Finally, in section~\ref{sec:conclusions} we summarize
the main results of the paper and draw conclusions.

\section{One-dimensional confined model}

\label{sec:model}

For simplicity and concreteness we consider the one-dimensional
Schr\"{o}dinger equation
\begin{equation}
-\frac{\hbar ^{2}}{2m}\psi ^{\prime \prime }(x)+V(x)\psi (x)=E\psi (x),
\label{eq:Schro}
\end{equation}
with the boundary conditions
\begin{equation}
\psi (L_{1})=\psi (L_{2})=0,  \label{eq:bc}
\end{equation}
that describes the quantum-mechanical behaviour of a particle of mass $m$
confined to a box with impenetrable walls at $x=L_{1}$ and $x=L_{2}$ under
the effect of a potential $V(x)$.

It is convenient to convert this equation into a dimensionless one by means
of the change of variables
\begin{equation}
q=\frac{x-x_{0}}{L},\;L=L_{2}-L_{1},\;L_{1}\leq x_{0}\leq L_{2},
\label{eq:q(x)}
\end{equation}
that leads to
\begin{equation}
-\varphi ^{\prime \prime }(q)+v(q)\varphi (q)=\epsilon \varphi (q),
\label{eq:Schro_2}
\end{equation}
where
\begin{eqnarray}
\varphi (q) &=&\psi \left( Lq+x_{0}\right) ,  \nonumber \\
v(q) &=&\frac{2mL^{2}}{\hbar ^{2}}V\left( Lq+x_{0}\right) ,  \nonumber \\
\epsilon &=&\frac{2mL^{2}}{\hbar ^{2}}E.  \label{eq:dimensionless_quant}
\end{eqnarray}
The boundary conditions now become
\begin{eqnarray}
\varphi (l_{1}) &=&\varphi (l_{2})=0,  \nonumber \\
l_{1} &=&\frac{L_{1}-x_{0}}{L},\;l_{2}=\frac{L_{2}-x_{0}}{L}.
\label{eq:bc_2}
\end{eqnarray}

\section{Power-series approaches}

\label{sec:approaches}

The rate of convergence of the approaches discussed in what follows may
depend on the choice of $x_{0}$; however, in order to keep present
discussion as simple as possible we arbitrarily choose $x_{0}=L_{1}$ so that
$l_{1}=0$ and $l_{2}=1$.

We assume that the potential can be expanded in a Taylor series about $q=0$:
\begin{equation}
v(q)=\sum_{j=0}^{\infty }v_{j}q^{j}.  \label{eq:v(q)_series}
\end{equation}
This fact allows us to try a power-series solution
\begin{equation}
\varphi (q)=\sum_{j=0}^{\infty }c_{j}q^{j},  \label{eq:psi_series}
\end{equation}
where $c_{0}=c_{2}=0$ because of the boundary condition at $q=0$ and the
differential equation.

If we substitute the series (\ref{eq:psi_series}) and (\ref{eq:v(q)_series})
for $\varphi (q)$ and $v(q)$ in equation~(\ref{eq:Schro_2}), respectively,
we obtain the recurrence relation
\begin{equation}
c_{j}=\frac{1}{j(j-1)}\left( \sum_{i=0}^{j-2}v_{j-i-2}c_{i}-\epsilon
c_{j-2}\right) ,\;j=3,4,\ldots .  \label{eq:C_j_rec_rel}
\end{equation}
All the coefficients are proportional to $c_{1}$ that we may arbitrarily
choose equal to unity. In this way we have: $c_{0}=0$, $c_{1}=1$, $c_{2}=0$
and the remaining $c_{j}$, $j>2$, are polynomial functions of the
dimensionless energy $\epsilon $.

In what follows we discuss three alternative approaches based on the
power-series ansatz
\begin{equation}
\varphi ^{[N]}(q)=\sum_{j=1}^{N}c_{j}q^{j}.  \label{eq:psi^N}
\end{equation}
The first one ($A_{1}$) consists of calculating the coefficients $c_{j}$, $%
j=3,4,\ldots ,N$ from the recurrence relation (\ref{eq:C_j_rec_rel}) and
then determining the approximate dimensionless energy from the boundary
condition at $q=1$:
\begin{equation}
\sum_{j=1}^{N}c_{j}(\epsilon )=0.  \label{eq:approx_bc}
\end{equation}
In this way we obtain an approximation $\epsilon (A_{1})$ to the eigenvalue.
This is exactly the approach used by Patil and Varshni\cite{PV09} for the
confined hydrogen atom.

In the second approach ($A_{2}$) we calculate the coefficients $c_{j}$, $%
j=3,4,\ldots ,N-1$, from the recurrence relation (\ref{eq:C_j_rec_rel}) and $%
c_{N}$ from the boundary condition (\ref{eq:approx_bc}) thus obtaining the
ansatz
\begin{equation}
\varphi ^{[N]}(q)=\sum_{j=1}^{N-1}c_{j}(\epsilon )\left( q^{j}-q^{N}\right) .
\label{eq:psi^N_2}
\end{equation}
Since this function satisfies both boundary conditions we can use it as a
trial function for the variational method and calculate the variational
integral
\begin{eqnarray}
W^{[N]}(\epsilon ) &=&\frac{\int_{0}^{1}\varphi ^{[N]}(q)H\varphi
^{[N]}(q)\,dq}{\int_{0}^{1}\varphi ^{[N]}(q)^{2}\,dq},  \nonumber \\
H &=&-\frac{d}{dq^{2}}+v(q),  \label{eq:W^N}
\end{eqnarray}
where $\epsilon $ plays the role of a variational parameter. The optimal
approximate eigenvalue is given by
\begin{eqnarray}
W^{[N]}(A_{2}) &=&W^{[N]}(\epsilon (A_{2}))  \nonumber \\
\left. \frac{dW^{[N]}(\epsilon )}{d\epsilon }\right| _{\epsilon =\epsilon
(A_{2})} &=&0.  \label{eq:W_variational}
\end{eqnarray}
In this case we obtain two estimates of the eigenvalue $\epsilon (A_{2})$
and $W^{[N]}(A_{2})$ that in general do not agree, and the variational
theorem tells us that $W^{[N]}(A_{2})$ is an upper bound to the exact
ground-state energy $\epsilon _{0}$.

As we will show in the next section, $\epsilon (A_{2})$ and $W^{[N]}(A_{2})$
approach each other as $N$ increases which suggests a third approach ($A_{3}$%
) that consists of obtaining the approximate eigenvalue $\epsilon (A_{3})$
as a root of the equation
\begin{equation}
\epsilon =W^{[N]}(\epsilon ).  \label{eq:epsilon=W}
\end{equation}

\section{Simple benchmark problem}

\label{sec:results}

In what follows we consider a simple model for the problem of a crystal in
an electric field. The equation for the semifree electron in a
one-dimensional crystal is given by (\ref{eq:Schro}) with $V(x)=Fex$, where $%
F$ is the strength of the electric field and $m$ and $-e$ are the electron
mass and charge, respectively\cite{RZ71}. When $L_{1}=0$ and $L_{2}=L$ the
corresponding dimensionless eigenvalue equation is given by (\ref{eq:Schro_2}%
) with $v(q)=\lambda q$, where $\lambda =2mL^{3}Fe/\hbar ^{2}$.

By means of the change of variables
\begin{equation}
z=\lambda ^{1/3}q-\frac{\epsilon }{\lambda ^{2/3}},
\end{equation}
the eigenvalue equation becomes the Airy equation\cite{AS72}
\begin{equation}
Y^{\prime \prime }(z)-zY(z)=0,
\end{equation}
where $Y(z)=\varphi \left( \lambda ^{-1/3}z+\lambda ^{-1}\epsilon \right) $.
The solution is a linear combination $Y(z)=C_{1}Ai(z)+C_{2}Bi(z)$ of the
Airy functions $Ai(z)$ and $Bi(z)$\cite{AS72}. The boundary conditions at $%
q=0$ and $q=1$ lead to the quantization condition
\begin{equation}
A_{i}\left( -\lambda ^{-2/3}\epsilon \right) B_{i}\left( \lambda
^{1/3}-\lambda ^{-2/3}\epsilon \right) -Ai\left( \lambda ^{1/3}-\lambda
^{-2/3}\epsilon \right) Bi\left( -\lambda ^{-2/3}\epsilon \right) =0,
\end{equation}
that yields $\epsilon (\lambda )$ for all $\lambda \neq 0$.

For $\lambda =0$ the exact solutions are those for the particle in a box
\begin{eqnarray}
\varphi _{n-1}(q) &=&\sqrt{2}\sin (n\pi q)  \nonumber \\
\epsilon _{n-1} &=&n^{2}\pi ^{2},\;n=1,2,\ldots
\end{eqnarray}

Tables \ref{tab:v=0} and \ref{tab:v=q} show results for $\lambda =0$ and $%
\lambda =1$, respectively. In both cases we appreciate that $W(A_{2})$ and $%
\epsilon (A_{3})$ are better estimates of $\epsilon _{0}$ than $\epsilon
(A_{1})$. Besides, $W(A_{2})$ is an upper bound to $\epsilon _{0}$ as
expected from its variational origin and $\epsilon (A_{2})$ and $W(A_{2})$
approach each other as $N$ increases. The exact results are $\epsilon
_{0}(\lambda =0)=\pi ^{2}\approx 9.8696044010893586191$ and $\epsilon
_{0}(\lambda =1)\approx 10.368507161836337127$.

Table~\ref{tab:A1} shows results for the approach $A_{1}$ for greater values
of $N$. While $A_{2}$ and $A_{3}$ yield an accuracy of 10 digits for $N=12$ (%
$\lambda =1$) and $N=13$ ($\lambda =0$), $A_{1}$ requires as much as $N=21$.
However, the latter approach is much simpler because it does not require the
calculation of integrals. The three approaches are based on a local
approximation (the Taylor series about $q=0$) but only $A_{1}$ is purely
local.

The variational approach $A_{2}$ is not the best one based on the expansion (%
\ref{eq:psi^N_2}). If we define the set of functions
\begin{equation}
f_{N,j}=q^{j}-q^{N},\;j=1,2,\ldots ,N-1  \label{eq:basis}
\end{equation}
then the Rayleigh-Ritz method leads to the secular equation\cite{P68}
\begin{equation}
\left( \mathbf{H}-\epsilon \mathbf{S}\right) \mathbf{C}=0,  \label{eq:sec_eq}
\end{equation}
where $\mathbf{H}$ and $\mathbf{S}$ are $(N-1)\times (N-1)$ square matrices
with elements
\begin{equation}
H_{ij}=\left\langle f_{N,i}\right| H\left| f_{N,j}\right\rangle
,\;S_{ij}=\left\langle f_{N,i}\right. \left| f_{N,j}\right\rangle ,
\label{eq:Hij-Sij}
\end{equation}
and $\mathbf{C}$ is a the column vector of the coefficients $c_{i}$, $%
i=1,2,\ldots ,N-1$. The approximate eigenvalues are given by the roots of
the secular determinant
\begin{equation}
|\mathbf{H}-\epsilon \mathbf{S|=0.}  \label{eq:sec_det}
\end{equation}
In order to apply the Rayleigh-Ritz method we have defined the scalar
product
\begin{equation}
\left\langle f\right. \left| g\right\rangle =\int_{0}^{1}f(q)g(q)\,dq.
\end{equation}

Table~\ref{tab:RR} shows approximate results for the lowest eigenvalue $%
\epsilon _{0}$. We appreciate that the rate of convergence of the
Rayleigh-Ritz method is greater than those for the approaches $A_{2}$ and $%
A_{3}$. However, we cannot consider that it is a true improvement of the
power-series method because it is based on an entirely different philosophy.

\section{Conclusions}

\label{sec:conclusions}

The purpose of this paper is to show that one can easily improve the rate of
convergence of the well-known power-series method for the calculation of
eigenvalues. The main idea is to introduce a global approximation into the
otherwise completely local approach. The main disadvantage of the global
approximations is that they require the calculation of integrals that one
may not be able to obtain analytically for some potentials $v(q)$. The use
of numerical integration in such unfavourable cases makes the global
approaches less appealing than the local one that completely avoids
integrals. The latter approach is just based on the Taylor expansion of the
potential-energy function about a chosen coordinate point.

We have also discussed the Rayleigh-Ritz variational method just for the
purpose of comparison. This approach is not based on the expansion about a
given point and is completely global despite the fact that the form of the
series (\ref{eq:psi^N_2}) led us to the choice of the basis set (\ref
{eq:basis}).

It is not difficult to extend the methods discussed above to more complex
problems. First, we choose a suitable ansatz $\psi (\mathbf{c},\mathbf{x})$,
where $\mathbf{c}$ is a set of linear parameters and $\mathbf{x}$ the set of
coordinates that describe the quantum-mechanical system. Second we force the
ansatz to satisfy the Schr\"{o}dinger equation $H\psi =E\psi $ and some of
its derivatives with respect to $\mathbf{x}$ at a chosen point $\mathbf{x}%
_{0} $. In this way we obtain $\mathbf{c}(\epsilon )$. Third, we
minimize the variational integral $W(\epsilon )$ with respect to
$\epsilon $ or solve the equation $\epsilon=W(\epsilon)$.

One can also imagine variants of the technique just outlined. For
example, we can force the ansatz to satisfy the Schr\"{o}dinger
equation at a set of coordinate points $\mathbf{x}_{i}$,
$i=0,1,\ldots $ (including some of the derivatives with respect to
the coordinates). Besides, the ansatz $\psi
(\mathbf{c},\mathbf{a},\mathbf{x})$ may also contain a set of
nonlinear adjustable parameters $\mathbf{a}$ that we obtain,
together with $\epsilon $, in the
variational part of the method that consists of minimizing $W(\mathbf{a}%
,\epsilon )$. As we appreciate, the method is quite flexible and offers
several variants that we may select in accordance with the chosen problem.

\begin{table}[H]
\caption{Estimates of $\epsilon_0$ for $\lambda=0$}
\label{tab:v=0}
\begin{center}
\par
\begin{tabular}{D{.}{.}{2}D{.}{.}{11}D{.}{.}{11}D{.}{.}{11}D{.}{.}{11}}
\hline \multicolumn{1}{c}{$N$}&
\multicolumn{1}{c}{$\epsilon(A_1)$} &
\multicolumn{1}{c}{$\epsilon(A_2)$} & \multicolumn{1}{c}{$W(A_2)$}& \multicolumn{1}{c}{$\epsilon(A_3)$} \\
\hline
 4 &       6       &  12.08941897   &   9.870757651 &  9.9717028   \\
 5 &     $--$        &   9.101852828  &   9.875388202 &  9.949871274 \\
 6 &     $--$        &   9.558639637  &   9.870985812 &  9.881622575 \\
 7 &   9.478038438 &   9.960092497  &   9.86992353  &  9.870713549 \\
 8 &   9.478038438 &   9.905617739  &   9.869662206 &  9.869825364 \\
 9 &   9.914249166 &   9.863621098  &   9.869607064 &  9.869612707 \\
10 &   9.914249166 &   9.867032812  &   9.869604943 &  9.869606146 \\
11 &   9.866812676 &   9.869907624  &   9.869604411 &  9.86960443  \\
12 &   9.866812676 &   9.869739577  &   9.869604403 &  9.869604407 \\
13 &   9.869737257 &   9.869592550  &   9.869604401 &  9.869604401 \\

\end{tabular}
\par
\end{center}
\end{table}

\begin{table}[H]
\caption{Estimates of $\epsilon_0$ for $\lambda=1$}
\label{tab:v=q}
\begin{center}
\par
\begin{tabular}{D{.}{.}{2}D{.}{.}{11}D{.}{.}{11}D{.}{.}{11}D{.}{.}{11}}
\hline \multicolumn{1}{c}{$N$}&
\multicolumn{1}{c}{$\epsilon(A_1)$} &
\multicolumn{1}{c}{$\epsilon(A_2)$} & \multicolumn{1}{c}{$W(A_2)$}& \multicolumn{1}{c}{$\epsilon(A_3)$} \\
\hline
 4 &     6.5         & 12.26637996   & 10.36907124  &10.44756346  \\
 5 &      $--$         &   9.434668415 &  10.37615238 & 10.50115579   \\
 6 &      $--$         &  10.11267971  &  10.36932531 & 10.37634015   \\
 7 &      9.568181651&  10.51534961  &  10.36928674 & 10.37131465   \\
 8 &     10          &  10.39424448  &  10.36853346 & 10.36861906   \\
 9 &     10.53613098 &  10.3551899   &  10.36851922 & 10.36854793   \\
10 &     10.40454178 &  10.36714011  &  10.36850729 & 10.36850764    \\
11 &     10.35308651 &  10.36945371  &  10.36850725 & 10.36850744 \\
12 &     10.36699657 &  10.36853572  &  10.36850716 & 10.36850716 \\
13 &     10.36959361 &  10.36845690  &  10.36850716 & 10.36850716  \\

\end{tabular}
\end{center}
\end{table}

\begin{table}[H]
\caption{Convergence of the estimate of $\epsilon_0$ from the method $A_1$}
\label{tab:A1}
\begin{center}
\par
\begin{tabular}{D{.}{.}{2}D{.}{.}{11}D{.}{.}{11}}
\hline \multicolumn{1}{c}{$N$}&
\multicolumn{1}{c}{$\epsilon(A_1,\lambda=0)$} &
\multicolumn{1}{c}{$\epsilon(A_1,\lambda=1)$}  \\
\hline

15 & 9.869599545   &    10.36845326 \\
16 & 9.869599545   &    10.36851066 \\
17 & 9.869604541   &    10.36850904 \\
18 & 9.869604541   &    10.36850690  \\
19 & 9.869604397   &    10.36850711 \\
20 & 9.869604397   &    10.36850717 \\
21 & 9.869604401   &    10.36850716 \\

\end{tabular}
\par
\end{center}
\end{table}

\begin{table}[tbp]
\caption{Rayleigh-Ritz method}
\label{tab:RR}
\begin{center}
\par
\begin{tabular}{D{.}{.}{2}D{.}{.}{11}D{.}{.}{11}}
\hline \multicolumn{1}{c}{$N$}& \multicolumn{1}{c}{$\lambda=0$} &
\multicolumn{1}{c}{$\lambda=1$}  \\
\hline

4 &   9.869749621  & 10.36873394  \\
6 &   9.869604434  & 10.36850740  \\
8 &   9.869604401  & 10.36850716  \\

\end{tabular}
\par
\end{center}
\end{table}


\begin{thebibliography}{9}
\bibitem{AQC09}  See, for example, Adv. Quantum Chem. volumes 57 and 58

\bibitem{PV09}  S. H. Patil and Y. P. Varshni, ''Properties of confined
hydrogen and helium atoms'', Adv. Quantum Chem. \textbf{57}, 1-24 (2009).

\bibitem{RZ71}  A. Rabinovitch and J. Zak, ''Electrons in crystals in a
finite-range electric field'', Phys. Rev. B \textbf{4}, 2358-2370 (1971).

\bibitem{AS72}  M. Abramowitz and I. A. Stegun, \textsl{Handbook of
Mathematical Functions}, Ninth ed. (Dover, New York, 1972).

\bibitem{P68}  F. L. Pilar, \textsl{Elementary Quantum Chemistry},
(McGraw-Hill, New York, 1968).
\end{thebibliography}
\end{document}